\documentclass[11pt,a4paper]{article}

\usepackage{amsmath}
\usepackage{amssymb}
\usepackage{amsthm}
\usepackage{authblk}
\usepackage{bussproofs}
\usepackage{cancel}
\usepackage[T1]{fontenc}
\usepackage{mathpartir}
\usepackage{styles/tr-coqvf}
\usepackage{styles/tr-listings}
\usepackage{tabularx}
\usepackage{tikz}
\usepackage{xcolor}

\usetikzlibrary{trees}

\usepackage{hyperref}

\newtheoremstyle{myplain}
  {\topsep}   
  {\topsep}   
  {}          
  {}          
  {\bfseries} 
  {.}         
  {5pt}       
  {}          

\theoremstyle{myplain}
\newtheorem{theorem}{Theorem}[section]
\newtheorem{lemma}[theorem]{Lemma}
\theoremstyle{definition}
\newtheorem{definition}[theorem]{Definition}

\newcommand{\vfcx}{VeriFast Cx}
\newcommand{\cbsem}{\texttt{cbsem}}

\newcommand{\coderef}[1]{\texttt{\detokenize{#1}}}

\newcommand{\margellipsis}{\llap{\textcolor{sol-base1}{$\vdots$}\hspace{-0.55cm}}}

\author[1]{Stefan Wils}
\author[2]{Bart Jacobs}
\affil[1,2]{IMEC-DistriNet, Computer Science Dept., KU Leuven\\Celestijnenlaan 200A, 3001 Leuven, Belgium}
\affil[1]{{\footnotesize{stefan.wils@kuleuven.be}}}

\title{Certifying C program correctness with respect to CH2O with VeriFast}

\begin{document}

\maketitle

\begin{abstract}
  VeriFast is a powerful tool for verification of various correctness properties of C programs using symbolic execution. However, VeriFast itself has not been verified. We present a proof-of-concept extension which generates a correctness certificate for each successful verification run individually. This certificate takes the form of a Coq script which, when successfully checked by Coq, removes the need for trusting in the correctness of VeriFast itself.

  The Coq script achieves this by applying a chain of soundness results, allowing us to prove correctness of the program with regards to the third-party CH2O small step semantics for C11 by proving correctness in terms of symbolic execution in Coq. This proof chain includes two intermediate auxiliary big step semantics, the most important of which describes VeriFast's interpretation of C. Finally, symbolic execution in Coq is implemented by transforming the exported AST of the program into a Coq proposition representing the symbolic execution performed by VeriFast itself.
\end{abstract}

\tableofcontents


\section{Introduction}

VeriFast \cite{jacobs-vf-tutorial-2017} is a general verification tool for C programs developed at KU Leuven. The input to VeriFast consists of a C program together with some annotations specifying, among other things, pre- and postconditions for each function, including the main function. VeriFast verifies the program using symbolic execution. A successful verification run implies that the program will not exhibit undefined behavior such as accessing unallocated memory or data races in any concrete execution.

VeriFast itself however is not formally verified. Aside from potential bugs, there is no formal account of how VeriFast's version of C relates to the ISO C standard. Closing this gap by \emph{direct} verification of VeriFast is difficult due to its internal complexity and rich set of features.

In our previous report \cite{wils-verifast-arxiv-2021}, we demonstrated how this problem can be solved by extending VeriFast to export a machine checkable proof upon a concrete instance of successful verification, allowing formal verification of \emph{that individual successful verification run}. This machine checkable proof takes the form of a Coq script that, when successfully type checked by the Coq compiler, proves that the C program verified by VeriFast is correct with respect to CompCert's \cite{leroy-compcert-cacm-2009} big step semantics for the Clight intermediate language \cite{blazy-clight-jar-2009}. Compiling this Coq script allows the user to conclude that the program will not exhibit undefined behavior under any concrete execution, based exclusively on trust in Coq's computational kernel and in CompCert's Clight semantics.

The present report serves as an update to our earlier work. In brief, we have made the following two improvements:
\begin{enumerate}
  \item We have completely reorganized our Coq code to use the STDPP utility library \cite{stdpp}. This has simplified many proofs by allowing us to use the powerful data structures, lemmas and solver tactics provided by this library.
  \item We have replaced the Clight backend with CH2O \cite{krebbers-ch2o-phd-2015}. CH2O is a formal semantics of C, faithful to the C11 standard. This means that our exported script proves correctness of the verified program \emph{for any compiler that is C11-compliant}. CH2O also features an interesting model of memory allowing both low-level operations (similar to the memory models of both VeriFast and Clight) but also compiler analyses such as type-based alias analysis based on a view of C as a high-level language.

  None of this means we have abandoned CompCert: since CompCert C is also C11-compliant, any program proven correct with respect to CH2O will also be correct with respect to CompCert C (see Section \ref{section:future_work} on future work for more details).
\end{enumerate}

To make this work, we had to refactor CH2O to use STDPP instead of its own internal ``precursor'' to STDPP. This was not only a \emph{nice to have}, but also necessary, because CH2O and STDPP both rely heavily on overriding the same notations, which caused conflicts. CH2O specifies two languages: \emph{abstract C}, which is close to C abstract syntax, and \emph{core C}, which is a simplified version of C for which the formal semantics is developed. Our soundness proof targets \emph{core C}, but we extended the CH2O command line tool to allow exports of parsed \emph{abstract C} code (this work has not yet been merged into the official CH2O release; it is however included in the Zenodo code drop described below).

\subsection{Overview}

Section \ref{section:example} demonstrates our approach for a simple example program, showing how the proof script exported by VeriFast is organized and how it employs our various soundness results to prove correctness of the verified program with respect to CH2O core C semantics.

Section \ref{section:vfcx} briefly recapitulates the work from the previous report to the extent necessary to make the new report independently readable. We describe the subset of C currently supported by our work (called \vfcx ), symbolic execution in Coq and our own coinductive big step semantics for \vfcx\ (called \cbsem ). We finish by describing the soundness result linking symbolic execution to \cbsem .

Section \ref{section:ch2o} presents our new results targeting CH2O: it introduces a \emph{second} coinductive big step semantics for CH2O core C and two new soundness results proving (1) soundness of \cbsem\ to this new CH2O big step semantics; and (2) soundness of this CH2O big step semantics to the actual CH2O small step semantics developed in \cite{krebbers-ch2o-phd-2015}.

We conclude this report with a brief discussion of related and future work in Section \ref{section:future_work}.

\subsection{Source code}

The source code for our experimental branch of VeriFast can be downloaded here:

\begin{center}
  \url{https://doi.org/10.5281/zenodo.7479859}
\end{center}

This code drop also includes a version of CH2O extended for exporting ASTs. Instructions for compiling VeriFast, together with all the required dependencies, are specified on the web page.

Our VeriFast extension not only \emph{exports} Coq code, but also includes a Coq \emph{library} containing the various definitions and theorems discussed in this report, together with the actual proofs. Instead of translating these definitions and theorems into a more compact symbolic form (as we did in the previous report), we will directly include the Coq code itself. The Coq code can be found under \coderef{src/coq} in the VeriFast repo.


\section{An example}
\label{section:example}

In this section we illustrate our approach to machine checkable verification runs using a concrete example program. This examples serves to show how the various stages of our approach, presented in more detail in the rest of the report, come together.

\subsection{Symbolic execution of a simple loop}
\label{subsection:example_program}

Listing \ref{lst:test_countdown.c} shows \coderef{tests/coq/test_countdown.c}, a simple C program implementing a countdown loop.

\vfclistinglabel{sources/test_countdown.c}{\coderef{tests/coq/test_countdown.c}}{lst:test_countdown.c}

\newcommand{\symexecbullet}{\ensuremath{\color{sol-orange} \bullet}}

\newcommand{\ASM}[1]{\ensuremath{\{ #1 \}}}
\newcommand{\ASMnew}[1]{\ensuremath{{\color{sol-green}#1}}}

\newcommand{\STR}[1]{\ensuremath{\langle[ #1 ]\rangle}}
\newcommand{\STRnew}[1]{\ensuremath{{\color{sol-green}#1}}}

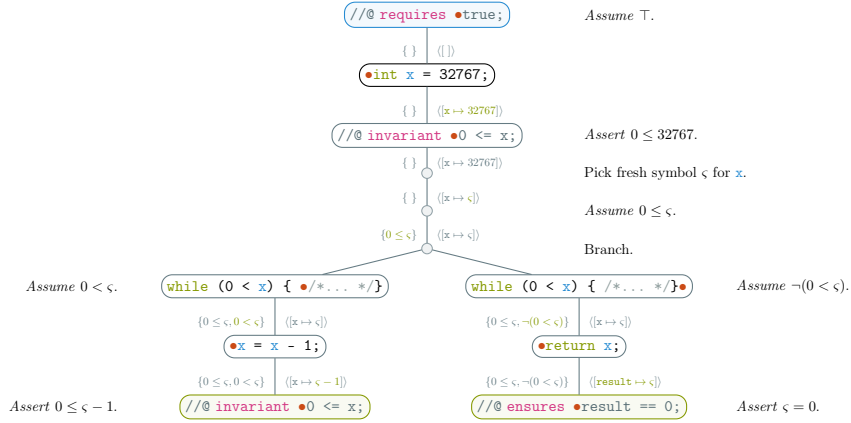
\begin{figure}[!ht]
  \centering
  \tikzstyle{symexec-node} = [draw]
  \tikzstyle{symexec-node-start} = [draw, color=sol-blue, fill=sol-blue!5]
  \tikzstyle{symexec-node-done} = [draw, color=sol-green, fill=sol-green!5]
  \tikzstyle{symexec-node-circ} = [circle, scale=0.7, draw, color=sol-base00!70, fill=sol-base00!10]
  \tikzstyle{symexec-edge} = [color=sol-base00]
  \tikzstyle{symexec-edge-l} = [left, scale=0.7, xshift=-0.2cm, yshift=-0.2cm, text=sol-base00]
  \tikzstyle{symexec-edge-r} = [right, scale=0.7, xshift=0.2cm, yshift=-0.2cm, text=sol-base00]
  \begin{tikzpicture}[
      every node/.style={text centered, rounded corners, scale=0.55},
      level 3/.style={level distance=0.5cm},
      level 7/.style={level distance=0.8cm},
      level distance=0.8cm,
      sibling distance=4cm
    ]
    \node (Root) [symexec-node-start] {\vfcinline{//@ requires $\ \symexecbullet$ true;}}
    child {
      node [symexec-node] {\vfcinline{$\symexecbullet$ int x = 32767;}}
      child {
        node [symexec-node] {\vfcinline{//@ invariant $\ \symexecbullet$ 0 <= x;}}
        child {
          node[symexec-node-circ] {}
          child {
            node[symexec-node-circ] {}
            child {
              node[symexec-node-circ] {}
              child {
                node [symexec-node] {\vfcinline{while (0 < x) \{ $\ \symexecbullet$ /* ... */ \}}}
                child {
                  node [symexec-node] {\vfcinline{$\symexecbullet$ x = x - 1;}}
                  child {
                    node  (LastL) [symexec-node-done] {\vfcinline{//@ invariant $\ \symexecbullet$ 0 <= x;}}
                    edge from parent [symexec-edge] {
                      node[symexec-edge-l] {$\ASM{ 0 \leq \varsigma, 0 < \varsigma }$}
                      node[symexec-edge-r] {$\STR{ \mathtt{x} \mapsto \STRnew{\varsigma - 1} }$}
                    }
                  }
                  edge from parent [symexec-edge]  {
                    node[symexec-edge-l] {$\ASM{ 0 \leq \varsigma, \ASMnew{0 < \varsigma} }$}
                    node[symexec-edge-r] {$\STR{ \mathtt{x} \mapsto \varsigma }$}
                  }
                }
                edge from parent [symexec-edge]
              }
              child {
                node [symexec-node] {\vfcinline{while (0 < x) \{ /* ... */ \} $\symexecbullet$}}
                child {
                  node [symexec-node] {\vfcinline{$\symexecbullet$ return x;}}
                  child {
                    node (LastR) [symexec-node-done] {\vfcinline{//@ ensures $\ \symexecbullet$ result == 0;}}
                    edge from parent [symexec-edge] {
                      node[symexec-edge-l] {$\ASM{ 0 \leq \varsigma, \neg (0 < \varsigma) }$}
                      node[symexec-edge-r] {$\STR{ \STRnew{\mathtt{result} \mapsto \varsigma} }$}
                    }
                  }
                  edge from parent [symexec-edge] {
                    node[symexec-edge-l] {$\ASM{ 0 \leq \varsigma, \ASMnew{\neg (0 < \varsigma)} }$}
                    node[symexec-edge-r] {$\STR{ \mathtt{x} \mapsto \varsigma }$}
                  }
                }
                edge from parent [symexec-edge]
              }
              edge from parent [symexec-edge] {
                node[symexec-edge-l] {$\ASM{ \ASMnew{0 \leq \varsigma} }$}
                node[symexec-edge-r] {$\STR{ \mathtt{x} \mapsto \varsigma }$}
              }
            }
            edge from parent [symexec-edge] {
              node[symexec-edge-l] {$\ASM{ \ }$}
              node[symexec-edge-r] {$\STR{ \mathtt{x} \mapsto \STRnew{\varsigma} }$}
            }
          }
          edge from parent [symexec-edge] {
            node[symexec-edge-l] {$\ASM{ \ }$}
            node[symexec-edge-r] {$\STR{ \mathtt{x} \mapsto 32767 }$}
          }
        }
        edge from parent [symexec-edge] {
          node[symexec-edge-l] {$\ASM{ \ }$}
          node[symexec-edge-r] {$\STR{ \STRnew{\mathtt{x} \mapsto 32767} }$}
        }
      }
      edge from parent [symexec-edge] {
        node[symexec-edge-l] {$\ASM{ \ }$}
        node[symexec-edge-r] {$\STR{ \ }$}
      }
    };

    \begin{scope}[
      every node/.style={right,scale=0.5}
    ]
      \path (Root           -| LastR) node {\emph{Assume} $\top$.};
      \path (Root-1-1       -| LastR) node {\emph{Assert} $0 \leq 32767$.};
      \path (Root-1-1-1     -| LastR) node {Pick fresh symbol $\varsigma$ for \vfcinline{x}.};
      \path (Root-1-1-1-1   -| LastR) node {\emph{Assume} $0 \leq \varsigma$.};
      \path (Root-1-1-1-1-1 -| LastR) node {Branch.};
    \end{scope}

    \begin{scope}[
      every node/.style={left,scale=0.5,xshift=-4cm}
    ]
      \path (Root-1-1-1-1-1-1     -| LastL) node {\emph{Assume} $0 < \varsigma$.};
      \path (LastL -| LastL) node {\emph{Assert} $0 \leq \varsigma - 1$.};
    \end{scope}

    \begin{scope}[
      every node/.style={right,scale=0.5,xshift=4cm}
    ]
      \path (Root-1-1-1-1-1-2     -| LastR) node {\emph{Assume} $\neg(0 < \varsigma)$.};
      \path (LastR -| LastR) node {\emph{Assert} $\varsigma = 0$.};
    \end{scope}
  \end{tikzpicture}
  \caption{A structural overview of symbolic execution in VeriFast.}
  \label{fig:symexec}
\end{figure}

Figure \ref{fig:symexec} shows in detail how VeriFast symbolically executes the above program. Each \emph{node} of this tree can be considered a breakpoint in the symbolic execution performed by VeriFast (marked by the red dot). On the left of each node is shown the current \emph{path condition} and on the right is shown the current \emph{symbolic store}. The symbolic store links each C variable to its current concrete or symbolic value. The path condition consists of a set of accumulated assumptions overestimating the current state of the symbolic execution. Each edge represents a state transition which can change the path condition and symbolic store. For clarity, changes to the path condition and to the symbolic store are rendered in green.

In the example, \vfcinline{main} is given the weakest possible precondition (\vfcinline{true} or $\top$), meaning nothing can be assumed initially. We then trace through the program, updating the path condition and symbolic store. The loop causes a branch in the symbolic execution: one branch verifies the loop body using an invariant and the other branch verifies the loop continuation. Since both branches satisfy their respective postconditions, VeriFast concludes that the function (and the entire program) is correct in terms of the C language and of its own specification.

\subsection{Certified verification of the example program}

We will now demonstrate our approach to certified program verification by applying it to the program shown in Listing \ref{lst:test_countdown.c}. (Again, instructions for setting up the VeriFast extension and its dependencies are provided on the download page of the Zenodo drop.)

\begin{enumerate}
  \item First, we need to run the extended \texttt{ch2o} command line tool on the program. This tool generates a Coq script (\coderef{test_ch2o.v}) containing the CH2O abstract C AST for the program:
  \begin{verbatim}$ ch2o -o tests/coq/test_countdown.c > test_ch2o.v\end{verbatim}

  \item Second, we run VeriFast on the C program using the new \texttt{-emit\_coq\_proof} flag:
  \begin{verbatim}$ bin/verifast -shared -emit_coq_proof -bindir bin
    tests/coq/test_countdown.c\end{verbatim}
  This generates a Coq script containing the VeriFast-generated correctness proof (\coderef{test_vf.v}). As we will see later, \coderef{test_vf.v} imports the \texttt{ch2o}-generated AST from \coderef{test_ch2o.v}.

  \item Finally, we can let Coq check our proof by compiling both scripts:
  \begin{verbatim}$ coqc test_ch2o.v
$ coqc test_vf.v -Q src/coq vf\end{verbatim}
\end{enumerate}
Let's inspect the contents of these scripts in more detail to understand what is going on.

\subsection{The Coq script exported by \texttt{ch2o}}
\label{subsection:coq_script_exported_by_ch2o}

When we run the \texttt{ch2o} tool on the above example, the resulting Coq script \coderef{test_ch2o.v} contains all the declarations of the program (in this case only function \vfcinline{main}):

\begin{lstlisting}[
  language=Coq,
  style=coqfootnotenumbered,
  linerange={7-22},
  firstnumber=7
]
Definition decls: list (string * decl) :=
  [("main",
    FunDecl [] (CTFun [] (CTInt (CIntType None CIntRank)))
      (CSLocal [] "x" (CTInt (CIntType None CIntRank))
        (Some
          (CSingleInit (CEConst (CIntType (Some Signed) CIntRank) 32767)))
        (CSComp
          (CSWhile
            (CEBinOp (CompOp LtOp)
              (CEConst (CIntType (Some Signed) CIntRank) 0) (CEVar "x"))
            (CSScope
              (CSDo
                (CEAssign Assign (CEVar "x")
                  (CEBinOp (ArithOp MinusOp) (CEVar "x")
                    (CEConst (CIntType (Some Signed) CIntRank) 1))))))
          (CSReturn (Some (CEVar "x"))))))].
\end{lstlisting}
As we will see next, it is this AST that will be imported from within the VeriFast export.

\subsection{The Coq script exported by VeriFast}

\subsubsection{Script header and architecture specification}
\label{subsection:example_script_header}

The second Coq script, the one exported by VeriFast, starts with some initial imports, followed by some boilerplate definitions related to the fact that the CH2O semantics can be parametrized by different architectures:
\begin{lstlisting}[
  language=Coq,
  style=coqfootnotenumbered,
  linerange={16-23},
  firstnumber=16
]
Definition A: architecture := {|
  arch_sizes := architectures.lp64;
  arch_big_endian := false;
  arch_char_signedness := integer_coding.Signed;
  arch_alloc_can_fail := false
|}.

Notation K := (arch_rank A).
\end{lstlisting}
For our proofs, we use the LP64 data type model.

\subsubsection{Program AST in \vfcx}

Next, we find the AST of the program \emph{as parsed by VeriFast}. Notably, this AST includes the VeriFast annotations, specifically the pre- and postcondition for \vfcinline{main} and the loop invariant.

Since we currently only support programs with a single main function taking no arguments, the data structure for a \vfcx\ program consists \emph{only} of the AST of that main function's body. The exported AST for the example program in \coderef{test_vf.v} looks like this:
\begin{lstlisting}[
  language=Coq,
  style=coqfootnotenumbered,
  linerange={40-61},
  firstnumber=40
]
Definition main_vf_func := simplify_vf_\sigmamt
  (StmtLet "x" ((ExprIntLit (32767))) (
    (StmtSeq
      (StmtWhile
        (ExprLt (ExprIntLit (0)) (ExprVar "x"))
        (ExprLe (ExprIntLit (0)) (ExprVar "x"))
        (StmtSeq
          (StmtBlock
            (StmtSeq
              (StmtExpr (ExprAssign (ExprVar "x") (ExprSub (ExprVar "x") (ExprIntLit (1)))))
              StmtSkip
            )
          )
          StmtSkip
        )
      )
      (StmtSeq
        (StmtRet (ExprVar "x"))
        StmtSkip
      )
    )
  )).
\end{lstlisting}
The function \coqinline{simplify_vf_\\sigmamt}, itself defined earlier within the exported script, simplifies AST sequences of the form \coqinline{(StmtSeq x StmtSkip)} to just \coqinline{x}. This makes it easier to line up \vfcx\ with CH2O.

\subsubsection{Importing and transforming the CH2O AST}

Following our definition of the \vfcx\ AST, we perform the actual import of the CH2O script from Subsection \ref{subsection:coq_script_exported_by_ch2o}:
\begin{lstlisting}[
  language=Coq,
  style=coqfootnotenumbered,
  linerange={65-65},
  firstnumber=65
]
Require test_ch2o.
\end{lstlisting}
The CH2O command-line tool exports the AST for \vfcinline{main} in CH2O abstract C, but our soundness results link VeriFast to CH2O core C. So, if we want to apply those soundness results in our correctness proof, we first need to \emph{translate} the CH2O abstract C AST for \vfcinline{main} into core C AST:
\begin{lstlisting}[
  language=Coq,
  style=coqfootnotenumbered,
  linerange={67-84},
  firstnumber=67
]
Definition alloc_program_result: frontend_state K.
  set (a:=alloc_program (K:=K) test_ch2o.decls \emptyset).
  assert (match a with inl _ => False | inr _ => True end). { apply I. }
  destruct a. { elim H. }
  destruct p.
  exact f.
Defined.

Definition \Gamma: env K := to_env alloc_program_result.
Definition \delta: funenv K := to_funenv alloc_program_result.
Definition m0: mem K := to_mem alloc_program_result.
Definition S0 := initial_state m0 "main" [].

Definition main_ch2o_func: stmt K.
  set (s := (funmap_lookup "main" \delta)).
  assert (match s with Some _ => True | None => False end). { apply I. }
  destruct s; done.
Defined.
\end{lstlisting}

\subsubsection{Establishing the relation between both ASTs}

Before we can get to the actual correctness, we first need to establish and prove a \emph{relation} between the core C and \vfcx\ ASTs for \vfcinline{main} (see Definition \ref{def:vfcx_cbsem__ch2o_cbsem.stmt_rel}):
\begin{lstlisting}[
  language=Coq,
  style=coqfootnotenumbered,
  linerange={88-91},
  firstnumber=88
]
Lemma rel: vfcx_cbsem__ch2o_cbsem.stmt_rel [] main_vf_func main_ch2o_func.
Proof.
  repeat (try constructor; try set_solver).
Qed.
\end{lstlisting}

\subsubsection{The actual correctness proof}
\label{subsubsection:the_actual_correctness_proof}

The actual correctness proof takes the form of a direct proof goal stating that the core C program \coqinline{main_ch2o_func} is correct in terms of CH2O core C's own small step operational semantics (see Definition \ref{def:ch2o_rs.exec_prog_correct}):
\begin{lstlisting}[
  language=Coq,
  style=coqfootnotenumbered,
  linerange={93-93},
  firstnumber=93
]
Goal prog_correct (K:=K) \GammaSPC \deltaSPC main_ch2o_func.
\end{lstlisting}
This goal is proven by applying our chain of soundness results \emph{in reverse}, reducing the goal to proving correctness in terms of VeriFast's symbolic execution. We begin by applying the soundness result from Theorem \ref{thm:ch2o_cbsem__ch2o_rs.exec_prog}, stating that the CH2O core C program is correct in terms of CH2O small step semantics if the same program is correct in terms of our own CH2O big step semantics:
\begin{lstlisting}[
  language=Coq,
  style=coqfootnotenumbered,
  linerange={94-94},
  firstnumber=94
]
  apply ch2o_cbsem__ch2o_rs.exec_prog.
\end{lstlisting}
Next, we apply the result from Theorem \ref{thm:vfcx_cbsem__ch2o_cbsem.exec_prog}, stating that the program in CH2O core C big step semantics is correct, if the corresponding \vfcx\ program is correct in \vfcx\ big step semantics (\cbsem ):
\begin{lstlisting}[
  language=Coq,
  style=coqfootnotenumbered,
  linerange={95-95},
  firstnumber=95
]
  apply (vfcx_cbsem__ch2o_cbsem.exec_prog _ _ rel).
\end{lstlisting}
Finally, we apply the soundness result from Theorem \ref{thm:vfcx_symexec__vfcx_cbsem.exec_prog_sound}, linking program correctness in \vfcx\ big step semantics to successful symbolic execution:
\begin{lstlisting}[
  language=Coq,
  style=coqfootnotenumbered,
  linerange={96-96},
  firstnumber=96
]
  apply vfcx_symexec__vfcx_cbsem.exec_prog.
\end{lstlisting}
After simplification using tactic \coqinline{cbv} (together with strategy guidelines to keep certain things opaque), we are left with proving a Coq \coqinline{Prop} called the ``symbolic execution proposition'' or SEP, which corresponds very closely to the tree shown earlier in \ref{fig:symexec}:
\begin{lstlisting}[
  language=Coq,
  style=coqfootnote,
]
True
\rightarrow (0 \leq 32767)%Z
  \land (\forall sym : Z,
       (-2147483648 \leq sym \leq 2147483647)%Z
       \rightarrow (0 \leq sym)%Z
         \rightarrow ((0 < sym)%Z
            \rightarrow (-2147483648 \leq sym - 1)%Z
              \land (sym - 1 \leq 2147483647)%Z \land (0 \leq sym - 1)%Z \land True)
           \land (((0 < sym)%Z \rightarrow False) \rightarrow True \land True))
\end{lstlisting}
This means that the final part of the proof consists of tactic applications discharging the proof obligations of the symbolic execution. The required tactics are recorded and exported by instrumenting VeriFast's symbolic execution engine:
\begin{lstlisting}[
  language=Coq,
  style=coqfootnotenumbered,
  linerange={99-125},
  firstnumber=99
]
  cbv.
  intros.  (* _: True *)
  split.
  (* assert 0 <= 32767 *)
  lia.
  intros.  (* x: Z *)
  intros.  (* _: 0 <= x *)
  split.
  intros.  (* _: 0 < x *)
  split.
  (* assert INT_MIN <= (x - 1) *)
  lia.
  split.
  (* assert (x - 1) <= INT_MAX *)
  lia.
  split.
  (* assert 0 <= (x - 1) *)
  lia.
  (* assert True *)
  lia.
  intros.  (* _: ~(0 < x) *)
  split.
  (* assert x = 0 *)
  lia.
  (* assert True *)
  lia.
Qed.
\end{lstlisting}
Given that the Coq compiler accepts this proof, we have just proven that this program meets its specification when interpreted in our \vfcx\ big-step semantics. And because Coq also accepts our soundness theorems, we have also proven that the program will show no undefined behavior as defined in the CH2O core C semantics.


\section{\vfcx : symbolic execution and big step semantics}
\label{section:vfcx}

In this section we very briefly reprise from \cite{wils-verifast-arxiv-2021} our description of \vfcx\ (the small C language subset currently supported by the export extension to VeriFast); our shallow embedding of VeriFast's symbolic execution for \vfcx\ programs; and the coinductive big step semantics for \vfcx\ programs.

\subsection{\vfcx}

\vfcx , while being a small subset of C, is still large enough to allow for programs that diverge and that can exhibit some straightforward forms of undefined behavior, specifically out-of-bounds arithmetic and division by zero. That being said, for now \vfcx\ misses some very relevant things such as supporting types other than \vfcinline{int}, function calls and heap memory.

\begin{definition}
  \label{def:vfcx.expr}

  The set of \vfcx\ \emph{expressions} currently supported by our export code is defined inductively in \coderef{vfcx.v} as:
\begin{lstlisting}[
    language=Coq,
    style=coqfootnotenumbered,
    linerange={11-27},
    firstnumber=11
  ]
Inductive expr :=
| ExprTrue
| ExprFalse
| ExprIntLit (z: Z)
| ExprVar (x: string)
| ExprAdd (e1: expr) (e2: expr)
| ExprSub (e1: expr) (e2: expr)
| ExprDiv (e1: expr) (e2: expr)
| ExprLt (e1: expr) (e2: expr)
| ExprLe (e1: expr) (e2: expr)
| ExprEq (e1: expr) (e2: expr)
| ExprNeq (e1: expr) (e2: expr)
| ExprAnd (e1: expr) (e2: expr)
| ExprOr (e1: expr) (e2: expr)
| ExprNot (e: expr)
| ExprAssign (e1: expr) (e2: expr)
.
\end{lstlisting}
  The set of expressions includes boolean comparison operators. Symbolic execution in Coq is implemented in such a way that these operators can only be used in certain places, such as the conditions of conditional statements and loops.
\end{definition}

\begin{definition}
  \label{def:vfcx.stmt}

  Likewise, the set of \vfcx\ \emph{statements} currently generated by our export code is defined inductively in \coderef{vfcx.v} as:
\begin{lstlisting}[
    language=Coq,
    style=coqfootnotenumbered,
    linerange={33-34},
    firstnumber=33,
    belowskip=-3pt
  ]
Inductive stmt :=
| StmtSkip
\end{lstlisting}
  \margellipsis
\begin{lstlisting}[
    language=Coq,
    style=coqfootnotenumbered,
    linerange={42-42},
    firstnumber=42,
    belowskip=-3pt,
    aboveskip=0pt
  ]
| StmtSeq (s1: stmt) (s2: stmt)
\end{lstlisting}
  \margellipsis
\begin{lstlisting}[
    language=Coq,
    style=coqfootnotenumbered,
    linerange={50-56},
    firstnumber=50,
    aboveskip=0pt
  ]
| StmtLet (x: string) (e: expr) (s: stmt)
| StmtExpr (e: expr)
| StmtIf (e: expr) (s1: stmt) (s2: stmt)
| StmtRet (e: expr)
| StmtWhile (e: expr) (i: expr) (s: stmt)
| StmtBlock (s: stmt)
.
\end{lstlisting}
\end{definition}

\subsection{Symbolic execution in Coq}
\label{subsection:symexec}

The main difference with our previous work described in \cite{wils-verifast-arxiv-2021} is that we now use STDPP's data structures.

\begin{definition}
  \label{def:store.store}
  Stores in our earlier version were represented as total functions of type \coqinline{string -> option Z}, whereas now they are represented using STDPP's \coqinline{stringmap Z}. From \coderef{store.v}:
\begin{lstlisting}[
    language=Coq,
    style=coqfootnotenumbered,
    linerange={13-13},
    firstnumber=13
  ]
Definition store := stringmap Z.
\end{lstlisting}
\end{definition}

Switching to STDPP required us to rewrite many low-level proofs but in the end has greatly simplified many of the proofs, because it allows us to rely on STDPP's abundance of utility lemmas and powerful tactics such as \coqinline{set_solver} to discharge goals.

VeriFast's symbolic execution of a function is implemented in Coq by \emph{shallow embedding}. The exported AST of the function, together with its pre- and postconditions, are fed into a Coq predicate expressing correct execution of a function.

\begin{definition}
  \label{def:vfcx_symexec.exec_func_correct}
  Correct symbolic execution of a \vfcx\ function is defined in \coderef{vfcx_symexec.v}:
\begin{lstlisting}[
    language=Coq,
    style=coqfootnotenumbered,
    linerange={806-814},
    firstnumber=806
  ]
Definition exec_func_correct xs p s q :=
  for_Zs xs (produce p (
    \lambdaSPC st_after_pre, (
      exec_stmt s
        (\lambdaSPC _, False)
        (ret st_after_pre (consume q leak_check))
        st_after_pre
    )
  )) \emptyset.
\end{lstlisting}
  \coqinline{for_Zs} picks fresh symbols the function arguments \coqinline{xs: stringset}; \coqinline{produce} provides an Coq embedding of the precondition \coqinline{p: expr}; fixpoint function \coqinline{exec_stmt} performs the bulk of the work of embedding symbolic execution for the function body \coqinline{s: stmt}; whereas \coqinline{consume} embeds the postcondition \coqinline{q: expr}.
\end{definition}

\begin{definition}
  \label{def:vfcx_symexec.exec_prog_correct}
  Correct symbolic execution of an entire \vfcx\ program is a trivial wrapper around the \coqinline{exec_func_correct} predicate:
\begin{lstlisting}[
    language=Coq,
    style=coqfootnotenumbered,
    linerange={816-816},
    firstnumber=816
  ]
Definition exec_prog_correct p s q := exec_func_correct \emptysetSPC p s q.
\end{lstlisting}
\end{definition}

The return value of this correctness predicate is a \coqinline{Prop} called the ``symbolic execution proposition'' (SEP), expressing the assumptions used and assertions made by VeriFast's SMT solver in the various branches of the symbolic execution. By application of Coq tactic \coqinline{cbv} (suitably constrained), the structure of this SEP becomes very close to the structure of the evaluation tree shown in Figure \ref{fig:symexec}.

As before, we have instrumented VeriFast to export a proof by recording calls to the internal SMT and \coderef{branch} function and use these recordings to generate a tactics proof.

\subsection{Big step semantics of \vfcx}

The final part of \cite{wils-verifast-arxiv-2021} that we are retaining, albeit also transformed by the adoption of STDPP, is the big step semantics for \vfcx . We started out with big step semantics because of their simplicity. To solve the issue of discriminating diverging programs from programs exhibiting undefined behavior, we specify two sets of semantics, one inductive for terminating programs and one coinductive for diverging programs. \cbsem\ refers to the \emph{combined} inductive and coinductive sets of semantic rules.

\begin{definition}
  \label{def:vfcx_cbsem.exec_stmt}

  The big step relation \coqinline{vfcx_cbsem.exec_stmt} describes the \emph{terminating execution} of some \vfcx\ statement \coqinline{s} relating some initial symbolic store \coqinline{\\sigma} with some final symbolic store \coqinline{\\sigma'} and outcome \coqinline{out}, which is either \coqinline{OutcomeNormal} or involves returning with some value (\coqinline{OutcomeReturn z}). It is inductively defined in \coderef{vfcx_cbsem.v}:
\begin{lstlisting}[
    language=Coq,
    style=coqfootnotenumbered,
    linerange={28-75},
    firstnumber=28
  ]
Inductive exec_stmt: store \rightarrowSPC stmt \rightarrowSPC store \rightarrowSPC outcome \rightarrowSPC Prop :=
| ExecSkip: \forallSPC \sigma,
    exec_stmt \sigmaSPC StmtSkip \sigmaSPC OutcomeNormal
| ExecSeq: \forallSPC \sigmaSPC \sigma' \sigma'' s1 s2 out,
    exec_stmt \sigmaSPC s1 \sigma' OutcomeNormal \rightarrow
    exec_stmt \sigma' s2 \sigma'' out \rightarrow
    exec_stmt \sigmaSPC (StmtSeq s1 s2) \sigma'' out
| ExecSeqAbnormal: \forallSPC \sigmaSPC \sigma' s1 s2 out,
    exec_stmt \sigmaSPC s1 \sigma' out \rightarrow
    out \neqSPC OutcomeNormal \rightarrow
    exec_stmt \sigmaSPC (StmtSeq s1 s2) \sigma' out
| ExecLet: \forallSPC \sigmaSPC \sigma' x e z s out,
    \sigmaSPC !! x = None \rightarrow
    eval_expr_to_Z e \sigmaSPC = Some z \rightarrow
    exec_stmt (<[ x := z ]> \sigma) s \sigma' out \rightarrow
    exec_stmt \sigmaSPC (StmtLet x e s) (delete x \sigma') out
| ExecExprAssign: \forallSPC \sigmaSPC x e z,
    is_Some (\sigmaSPC !! x) \rightarrow
    eval_expr_to_Z e \sigmaSPC = Some z \rightarrow
    exec_stmt \sigmaSPC (StmtExpr (ExprAssign (ExprVar x) e)) (<[ x := z ]> \sigma) OutcomeNormal
| ExecIfTrue: \forallSPC \sigmaSPC \sigma' e s1 s2 out1,
    eval_expr_to_bool e \sigmaSPC = Some true \rightarrow
    exec_stmt \sigmaSPC s1 \sigma' out1 \rightarrow
    exec_stmt \sigmaSPC (StmtIf e s1 s2) \sigma' out1
| ExecIfFalse: \forallSPC \sigmaSPC \sigma' e s1 s2 out2,
    eval_expr_to_bool e \sigmaSPC = Some false \rightarrow
    exec_stmt \sigmaSPC s2 \sigma' out2 \rightarrow
    exec_stmt \sigmaSPC (StmtIf e s1 s2) \sigma' out2
| ExecReturn: \forallSPC \sigmaSPC e z,
    eval_expr_to_Z e \sigmaSPC = Some z \rightarrow
    exec_stmt \sigmaSPC (StmtRet e) \sigmaSPC (OutcomeReturn z)
| ExecWhileTrueAbnormal: \forallSPC \sigmaSPC \sigma' e i s out,
    eval_expr_to_bool e \sigmaSPC = Some true \rightarrow
    exec_stmt \sigmaSPC s \sigma' out \rightarrow
    out \neqSPC OutcomeNormal \rightarrow
    exec_stmt \sigmaSPC (StmtWhile e i s) \sigma' out
| ExecWhileTrueNormal: \forallSPC \sigmaSPC \sigma' \sigma'' e i s out,
    eval_expr_to_bool e \sigmaSPC = Some true \rightarrow
    exec_stmt \sigmaSPC s \sigma' OutcomeNormal \rightarrow
    exec_stmt \sigma' (StmtWhile e i s) \sigma'' out \rightarrow
    exec_stmt \sigmaSPC (StmtWhile e i s) \sigma'' out
| ExecWhileFalse: \forallSPC \sigmaSPC e i s,
    eval_expr_to_bool e \sigmaSPC = Some false \rightarrow
    exec_stmt \sigmaSPC (StmtWhile e i s) \sigmaSPC OutcomeNormal
| ExecBlock: \forallSPC \sigmaSPC \sigma' s out,
    exec_stmt \sigmaSPC s \sigma' out \rightarrow
    exec_stmt \sigmaSPC (StmtBlock s) \sigma' out
.
\end{lstlisting}
\end{definition}

\begin{definition}
  \label{def:vfcx_cbsem.execinf_stmt}

  The big step relation \coqinline{vfcx_cbsem.execinf_stmt} describes the
  \emph{diverging execution} of statement \coqinline{s} starting from some store \coqinline{\\sigma}. Since it does not terminate, it is not associated with a final state or an outcome. It is coinductively defined in \coderef{vfcx_cbsem.v}:
\begin{lstlisting}[
    language=Coq,
    style=coqfootnotenumbered,
    linerange={221-254},
    firstnumber=221
  ]
CoInductive execinf_stmt: store \rightarrowSPC stmt \rightarrowSPC Prop :=
| ExecinfSeq1: \forallSPC \sigmaSPC s1 s2,
    execinf_stmt \sigmaSPC s1 \rightarrow
    execinf_stmt \sigmaSPC (StmtSeq s1 s2)
| ExecinfSeq2: \forallSPC \sigmaSPC \sigma' s1 s2,
    exec_stmt \sigmaSPC s1 \sigma' OutcomeNormal \rightarrow
    execinf_stmt \sigma' s2 \rightarrow
    execinf_stmt \sigmaSPC (StmtSeq s1 s2)
| ExecinfLet: \forallSPC \sigmaSPC x e z s,
    \sigmaSPC !! x = None \rightarrow
    eval_expr_to_Z e \sigmaSPC = Some z \rightarrow
    execinf_stmt (<[ x := z ]> \sigma) s \rightarrow
    execinf_stmt \sigmaSPC (StmtLet x e s)
| ExecinfIfTrue: \forallSPC \sigmaSPC e s1 s2,
    eval_expr_to_bool e \sigmaSPC = Some true \rightarrow
    execinf_stmt \sigmaSPC s1 \rightarrow
    execinf_stmt \sigmaSPC (StmtIf e s1 s2)
| ExecinfIfFalse: \forallSPC \sigmaSPC e s1 s2,
    eval_expr_to_bool e \sigmaSPC = Some false \rightarrow
    execinf_stmt \sigmaSPC s2 \rightarrow
    execinf_stmt \sigmaSPC (StmtIf e s1 s2)
| ExecinfWhileBody: \forallSPC \sigmaSPC e i s,
    eval_expr_to_bool e \sigmaSPC = Some true \rightarrow
    execinf_stmt \sigmaSPC s \rightarrow
    execinf_stmt \sigmaSPC (StmtWhile e i s)
| ExecinfWhileLoop: \forallSPC \sigmaSPC \sigma' e i s,
    eval_expr_to_bool e \sigmaSPC = Some true \rightarrow
    exec_stmt \sigmaSPC s \sigma' OutcomeNormal \rightarrow
    execinf_stmt \sigma' (StmtWhile e i s) \rightarrow
    execinf_stmt \sigmaSPC (StmtWhile e i s)
| ExecinfBlock: \forallSPC \sigmaSPC s,
    execinf_stmt \sigmaSPC s \rightarrow
    execinf_stmt \sigmaSPC (StmtBlock s)
.
\end{lstlisting}
\end{definition}

\begin{definition}
  \label{def:vfcx_cbsem.exec_func_correct}
  The predicate describing correctness of a function in \cbsem\ can be found in \coderef{vfcx_cbsem.v}:
\begin{lstlisting}[
    language=Coq,
    style=coqfootnotenumbered,
    linerange={258-273},
    firstnumber=258
  ]
Definition exec_func_correct xs p s q: Prop :=
  \forallSPC \sigma,
    wb \sigmaSPC \rightarrow
    store_equiv_mod xs \emptysetSPC \sigmaSPC \rightarrow
    xs \subseteqSPC dom stringset \sigmaSPC \rightarrow
    translate_expr_to_Prop_cps p (\lambdaSPC P,
      P \rightarrow
      (
        \existsSPC \sigma' v,
          translate_expr_to_Prop_cps q (\lambdaSPC Q,
            exec_stmt \sigmaSPC s \sigma' (OutcomeReturn v) \landSPC Q
          ) (<[ "result" := v ]> \sigma)
      ) \lorSPC (
        execinf_stmt \sigmaSPC s
      )
    ) \sigma.
\end{lstlisting}
  The definition states that a function is correct if, given that the precondition \coqinline{p} (embedded as a Coq \coqinline{Prop}) holds, then:
  \begin{enumerate}
    \item either we can construct an inductive derivation using \coqinline{exec_stmt} from Definition \ref{def:vfcx_cbsem.exec_stmt}, showing that the function body \coqinline{s} terminates with some final store \coqinline{\\sigmaprime} by returning some value \coqinline{v}. In addition, the postcondition \coqinline{q} must hold when given access to the return value of the execution through \vfcx\ variable \vfcinline{result}).
    \item or we can construct a coinductive derivation using \coqinline{execinf_stmt} from Definition \ref{def:vfcx_cbsem.execinf_stmt}, showing that body \coqinline{s} starting from \coqinline{\\sigma} diverges.
  \end{enumerate}
  The functional correctness definition has some relevant constraints on the initial store \coqinline{\\sigma}:
  \begin{enumerate}
    \item \coqinline{wb \\sigma}: all values stored in \coqinline{\\sigma} are within the limits of the integer type;
    \item \coqinline{store_equiv_mod xs \\emptyset \\sigma}: the store is equivalent to the empty store except for the function arguments \coqinline{xs};
    \item \coqinline{xs \\subseteq dom stringset \\sigma}: the function arguments are a subset of the domain of the store.
  \end{enumerate}
\end{definition}
\begin{definition}
  \label{def:vfcx_cbsem.exec_prog_correct}
  Correctness of an entire program is trivial. From \coderef{vfcx_cbsem.v}:
\begin{lstlisting}[
    language=Coq,
    style=coqfootnotenumbered,
    linerange={277-277},
    firstnumber=277
  ]
Definition exec_prog_correct p s q: Prop := exec_func_correct \emptysetSPC p s q.
\end{lstlisting}
\end{definition}

One of the main results of \cite{wils-verifast-arxiv-2021} was the soundness of symbolic execution described in Definitions \ref{def:vfcx_cbsem.exec_func_correct} and \ref{def:vfcx_cbsem.exec_prog_correct} with respect to \cbsem.

\begin{theorem}[Soundness of symbolic execution of a function with respect to \cbsem]
  \label{thm:vfcx_symexec__vfcx_cbsem.exec_func_sound}
  The theorem and its proof can be found in \coderef{vfcx_symexec__vfcx_cbsem.v}:
\begin{lstlisting}[
    language=Coq,
    style=coqfootnotenumbered,
    linerange={316-318},
    firstnumber=316
  ]
Theorem symexec_func_sound: \forallSPC xs p s q,
  vfcx_symexec.exec_func_correct xs p s q \rightarrow
  vfcx_cbsem.exec_func_correct xs p s q.
\end{lstlisting}
  It states that when correctness of a function has been proven by symbolic execution (that is, by proving the SEP constructed using Definition \ref{def:vfcx_symexec.exec_func_correct}), we may conclude correctness of a function as stated in Definition \ref{def:vfcx_cbsem.exec_func_correct}.
\end{theorem}

\begin{theorem}[Soundness of symbolic execution of a program with respect to \cbsem]
  \label{thm:vfcx_symexec__vfcx_cbsem.exec_prog_sound}
  Given the simplicity of language features currently supported, the soundness for entire programs follows trivially from the soundness for functions:
\begin{lstlisting}[
    language=Coq,
    style=coqfootnotenumbered,
    linerange={349-351},
    firstnumber=349
  ]
Theorem exec_prog: \forallSPC p s\_vSPC q,
  vfcx_symexec.exec_prog_correct p s\_vSPC q \rightarrow
  vfcx_cbsem.exec_prog_correct p s\_vSPC q.
\end{lstlisting}
\end{theorem}


\section{Soundness with respect to CH2O core C}
\label{section:ch2o}

\subsection{Big step semantics for CH2O core C}

One difference between \vfcx\ and CH2O core C is that core C has a single generic loop construct (\coqinline{loop s}) with flow control being implemented using a \coqinline{throw}/\coqinline{catch} mechanism. Another difference between \vfcx\ and CH2O core C is that \vfcx\ has a concept of non-addressable \emph{temporaries}: local variables that are taken from a store and of which the address cannot be taken. CH2O core C however stores local variables in memory. In addition, \vfcx\ doesn't allow uninitialized variables.

Instead of proving the soundness of \cbsem\ \emph{directly} with regards to the small step semantics included with CH2O, we decided to introduce a second coinductive big step semantics, one for CH2O core C, that can serve as an intermediary. This approach allows us to separate concerns associated with going from the \vfcx\ language to the CH2O core C language from the concerns associated with going from big step semantics to small step semantics and CH2O's lack of temporaries. In order to get started, we chose the subset of CH2O core C required to deal with our example programs.

\begin{definition}
  \label{def:ch2o_store.store}
  Given that CH2O core C uses de Bruijn indices, we define the notion of a CH2O big step store as a simple list. From \coderef{ch2o_store.v}:
\begin{lstlisting}[
    language=Coq,
    style=coqfootnotenumbered,
    linerange={7-7},
    firstnumber=7
  ]
Definition store := list (option Z).
\end{lstlisting}
\end{definition}

In the rest of this subsection, types \coqinline{expr K} and \coqinline{stmt K} (parametrized with an architecture rank \coqinline{K}, as can be seen in the example in subsection \ref{subsection:example_script_header}) refer to expressions and statements of CH2O core C. The specifications of these inductive types can be found in Definitions 6.1.4 and 6.2.2 of Krebbers's thesis \cite{krebbers-ch2o-phd-2015}, respectively. We will directly use the shorthand notation introduced by the CH2O Coq code, assuming it clear from the context by default and clarifying it where necessary. To distinguish between \vfcx\ and CH2O core C statements, expressions and stores, we will occasionally use subscripts \coqinline{v} (e.g. \coqinline{s\\_v}, \coqinline{e\\_v} or \coqinline{\\sigma\\_v}) for \vfcx\ and \coqinline{h} (e.g. \coqinline{s\\_h}, \coqinline{e\\_h} or \coqinline{\\sigma\\_h}) for CH2O-related variables.

\begin{definition}
  \label{def:ch2o_cbsem.eval}
  Expression evaluation for the subset of core C expressions of type \coqinline{expr K} is defined inductively in \coderef{ch2o_cbsem.v}:
\begin{lstlisting}[
    language=Coq,
    style=coqfootnotenumbered,
    linerange={41-44},
    firstnumber=41
  ]
Inductive eval `{Env K}: store \rightarrowSPC expr K \rightarrowSPC Z \rightarrowSPC Prop :=
| EvalLit: \forallSPC \sigmaSPC z,
    int_typed z sintT \rightarrow
    eval \sigmaSPC (# intV{sintT} z) z
\end{lstlisting}
  Using a local variable in an expression in CH2O big step means that its value must be present in a store as described in Definition \ref{def:ch2o_store.store}:
\begin{lstlisting}[
    language=Coq,
    style=coqfootnotenumbered,
    linerange={45-47},
    firstnumber=45
  ]
| EvalLoadVar: \forallSPC \sigmaSPC i z,
    \sigmaSPC !! i = Some (Some z) \rightarrow
    eval \sigmaSPC (load (var i)) z
\end{lstlisting}
  The unary and binary operations currently supported correspond to the operations we currently support in \vfcx:
\begin{lstlisting}[
    language=Coq,
    style=coqfootnotenumbered,
    linerange={48-59},
    firstnumber=48
  ]
| EvalUnop: \forallSPC \sigmaSPC e z op,
    eval \sigmaSPC e z \rightarrow
    unop_ok op z \rightarrow
    int_typed (unop_value op z) sintT \rightarrow
    eval \sigmaSPC (.{ op } e) (unop_value op z)
| EvalBinop: \forallSPC \sigmaSPC e1 z1 e2 z2 op,
    eval \sigmaSPC e1 z1 \rightarrow
    eval \sigmaSPC e2 z2 \rightarrow
    binop_ok op z1 z2 \rightarrow
    int_typed (binop_value op z1 z2) sintT \rightarrow
    eval \sigmaSPC (e1 .{ op } e2) (binop_value op z1 z2)
.
\end{lstlisting}
  Predicates \coqinline{unop_ok} and \coqinline{binop_ok} are defined inductively and perform the actual work of limiting which operations are supported.
\end{definition}

As with the coinductive big semantics for \vfcx\ described in Definitions \ref{def:vfcx_cbsem.exec_stmt} and \ref{def:vfcx_cbsem.execinf_stmt}, we also introduce two separate sets of big step rules for CH2O.

\begin{definition}
  \label{def:ch2o_cbsem.exec_stmt}
  The relation \coqinline{ch2o_cbsem.exec_stmt} describes the \emph{terminating execution} of a CH2O core C statement of type \coqinline{stmt K} starting from a CH2O store \coqinline{\\sigma} and generating one of the following three outcomes:
  \begin{enumerate}
    \item a \emph{normal} outcome (\coqinline{ONormal}), associated with a final store \coqinline{\\sigma'} (this is different from normal outcomes in \cbsem );
    \item a \emph{returning} outcome (\coqinline{OReturn}) associated with some return value;
    \item a \emph{throwing} outcome (\coqinline{OThrow}) associated with throw block counter and a final state, which is related to the way in which CH2O core C implements C controls structures.
  \end{enumerate}
  The relation is defined inductively in \coderef{ch2o_cbsem.v}:
\begin{lstlisting}[
    language=Coq,
    style=coqfootnotenumbered,
    linerange={84-140},
    firstnumber=84
  ]
Inductive exec_stmt `{Env K}: store \rightarrowSPC stmt K \rightarrowSPC outcome \rightarrowSPC Prop :=
| ExecSkip \sigma:
    exec_stmt \sigmaSPC skip (ONormal \sigma)
| ExecComp \sigmaSPC s1 \sigma' s2 O:
    exec_stmt \sigmaSPC s1 (ONormal \sigma') \rightarrow
    exec_stmt \sigma' s2 O \rightarrow
    exec_stmt \sigmaSPC (s1 ;; s2) O
| ExecCompAbnormal \sigmaSPC s1 s2 O:
    abnormal O \rightarrow
    exec_stmt \sigmaSPC s1 O \rightarrow
    exec_stmt \sigmaSPC (s1 ;; s2) O
| ExecLocal \sigmaSPC s mv \sigma':
    exec_stmt (None::\sigma) s (ONormal (mv::\sigma')) \rightarrow
    exec_stmt \sigmaSPC (local{sintT%BT} s) (ONormal \sigma')
| ExecLocalReturn \sigmaSPC s z:
    exec_stmt (None::\sigma) s (OReturn z) \rightarrow
    exec_stmt \sigmaSPC (local{sintT%BT} s) (OReturn z)
| ExecLocalThrow \sigmaSPC s n mv \sigma':
    exec_stmt (None::\sigma) s (OThrow n (mv::\sigma')) \rightarrow
    exec_stmt \sigmaSPC (local{sintT%BT} s) (OThrow n \sigma')
| ExecAssign \sigmaSPC i e z:
    i < length \sigmaSPC \rightarrow
    eval \sigmaSPC e z \rightarrow
    exec_stmt \sigmaSPC (var i ::= cast{sintT%BT} e) (ONormal (<[i:=Some z]>\sigma))
| ExecAssign' \sigmaSPC i e z:
    i < length \sigmaSPC \rightarrow
    eval \sigmaSPC e z \rightarrow
    exec_stmt \sigmaSPC (! (cast{voidT%T} (var i ::= e))) (ONormal (<[i:=Some z]>\sigma))
| ExecIf \sigmaSPC e z s1 s2 O:
    eval \sigmaSPC e z \rightarrow
    exec_stmt \sigmaSPC (if Z.eqb z 0 then s2 else s1) O \rightarrow
    exec_stmt \sigmaSPC (if{e} s1 else s2) O
| ExecThrow \sigmaSPC n:
    exec_stmt \sigmaSPC (throw n) (OThrow n \sigma)
| ExecCatch \sigmaSPC s \sigma':
    exec_stmt \sigmaSPC s (OThrow 0 \sigma') \rightarrow
    exec_stmt \sigmaSPC (catch s) (ONormal \sigma')
| ExecCatchSn \sigmaSPC s n \sigma':
    exec_stmt \sigmaSPC s (OThrow (S n) \sigma') \rightarrow
    exec_stmt \sigmaSPC (catch s) (OThrow n \sigma')
| ExecCatchNothrown \sigmaSPC s O:
    nothrown O \rightarrow
    exec_stmt \sigmaSPC s O \rightarrow
    exec_stmt \sigmaSPC (catch s) O
| ExecLoop \sigmaSPC s O:
    abnormal O \rightarrow
    exec_stmt \sigmaSPC s O \rightarrow
    exec_stmt \sigmaSPC (loop s) O
| ExecLoopIter \sigmaSPC s O \sigma':
    abnormal O \rightarrow
    exec_stmt \sigmaSPC s (ONormal \sigma') \rightarrow
    exec_stmt \sigma' (loop s) O \rightarrow
    exec_stmt \sigmaSPC (loop s) O
| ExecRet \sigmaSPC e z:
    eval \sigmaSPC e z \rightarrow
    exec_stmt \sigmaSPC (ret (cast{sintT%BT} e)) (OReturn z)
.
\end{lstlisting}
\end{definition}

\begin{definition}
  \label{def:ch2o_cbsem.execinf_stmt}
  The relation \coqinline{ch2o_cbsem.execinf_stmt} coinductively describes the \emph{diverging execution} of a CH2O core C statement of type \coqinline{stmt K} starting from a CH2O store \coqinline{\\sigma}. Also from \coderef{ch2o_cbsem.v}:
\begin{lstlisting}[
    language=Coq,
    style=coqfootnotenumbered,
    linerange={203-228},
    firstnumber=203
  ]
CoInductive execinf_stmt `{Env K}: store \rightarrowSPC stmt K \rightarrowSPC Prop :=
| ExecinfComp1 \sigmaSPC s1 s2:
    execinf_stmt \sigmaSPC s1 \rightarrow
    execinf_stmt \sigmaSPC (s1 ;; s2)
| ExecinfComp2 \sigmaSPC s1 \sigma' s2:
    exec_stmt \sigmaSPC s1 (ONormal \sigma') \rightarrow
    execinf_stmt \sigma' s2 \rightarrow
    execinf_stmt \sigmaSPC (s1 ;; s2)
| ExecinfLocal \sigmaSPC s:
    execinf_stmt (None::\sigma) s \rightarrow
    execinf_stmt \sigmaSPC (local{sintT%BT} s)
| ExecinfIf \sigmaSPC e z s1 s2:
    eval \sigmaSPC e z \rightarrow
    execinf_stmt \sigmaSPC (if Z.eqb z 0 then s2 else s1) \rightarrow
    execinf_stmt \sigmaSPC (if{e} s1 else s2)
| ExecinfCatch \sigmaSPC s:
    execinf_stmt \sigmaSPC s \rightarrow
    execinf_stmt \sigmaSPC (catch s)
| ExecinfLoop \sigmaSPC s:
    execinf_stmt \sigmaSPC s \rightarrow
    execinf_stmt \sigmaSPC (loop s)
| ExecinfLoopIter \sigmaSPC \sigma' s:
    exec_stmt \sigmaSPC s (ONormal \sigma') \rightarrow
    execinf_stmt \sigma' (loop s) \rightarrow
    execinf_stmt \sigmaSPC (loop s)
.
\end{lstlisting}
\end{definition}

\begin{definition}
  \label{def:ch2o_cbsem.exec_func_correct}
  The predicate describing correct execution of a CH2O core C function is structurally similar to the corresponding predicate for \cbsem\ from Definition \ref{def:vfcx_cbsem.exec_func_correct}, in the sense that it essentially consists of a disjunction of terminating and divergent execution. From \coderef{ch2o_cbsem.v}:
\begin{lstlisting}[
    language=Coq,
    style=coqfootnotenumbered,
    linerange={232-238},
    firstnumber=232
  ]
Definition exec_func_correct `{Env K} n s\_hSPC :=
  \lambdaSPC \sigma\_h,
    base.length \sigma\_hSPC = n \rightarrow
    store_typed \sigma\_hSPC \rightarrow
    (\existsSPC z, exec_stmt \sigma\_hSPC s\_hSPC (OReturn z))
    \lor
    execinf_stmt \sigma\_hSPC s\_h.
\end{lstlisting}
  \coqinline{n: nat} specifies the size of the CH2O store. Predicate \coqinline{store_typed} plays the same role as predicate \coqinline{wb} mentioned in Definition \ref{def:vfcx_cbsem.exec_func_correct}, ensuring that all integer values in the local store are within the limits of their type.
\end{definition}

\begin{definition}
  \label{def:ch2o_cbsem.exec_prog_correct}
  Program correctness is again a trivial wrapper around function correctness:
\begin{lstlisting}[
    language=Coq,
    style=coqfootnotenumbered,
    linerange={242-242},
    firstnumber=242
  ]
Definition exec_prog_correct `{Env K} s\_hSPC := exec_func_correct 0 s\_hSPC [].
\end{lstlisting}
\end{definition}

\subsection{Soundness of \cbsem\ with regards to CH2O big step semantics}

In order to construct a soundness proof of \cbsem\ with respect to CH2O core C big step semantics, we need to keep track of updates to their respective store implementations.

\begin{definition}
  The relation between a \vfcx\ store (Definition \ref{def:store.store}) and a CH2O store (Definition \ref{def:ch2o_store.store}) is done by relation \coqinline{store_rel}, which is defined inductively in \coderef{vfcx_cbsem__ch2o_cbsem.v}:
\begin{lstlisting}[
    language=Coq,
    style=coqfootnotenumbered,
    linerange={76-87},
    firstnumber=76
  ]
Inductive store_rel: list string \rightarrowSPC vf.store.store \rightarrowSPC vf.ch2o.ch2o_store.store \rightarrowSPC Prop :=
| store_rel_empty:
    store_rel [] \emptysetSPC []
| store_rel_push: \forallSPC xs \sigma\_vSPC x \sigma\_h,
    x \notinSPC xs \rightarrow
    store_rel xs \sigma\_vSPC \sigma\_hSPC \rightarrow
    store_rel (x :: xs) \sigma\_vSPC (None :: \sigma\_h)
| store_rel_alter: \forallSPC x xs \sigma\_vSPC \sigma\_hSPC i v,
    xs !! i = Some x \rightarrowSPC  (* same as: x \inSPC xs *)
    store_rel xs \sigma\_vSPC \sigma\_hSPC \rightarrow
    store_rel xs (<[x:=v]> \sigma\_v) (<[i:=Some v]> \sigma\_h)
.
\end{lstlisting}
  Constructor \coqinline{store_rel_push} illustrates the fact that \vfcx\ currently does not support uninitialized variables. This is why we separately keep track of all declared local variables in a list.
\end{definition}

We also need to map \vfcx\ programs onto CH2O core C programs. This is the task of three further relations: \coqinline{expr_rel_int} (for expressions involving arithmetic operations), \coqinline{expr_rel_bool} (for expressions involving logical operations) and \coqinline{stmt_rel}. For this report, we only explicitly mention \coqinline{stmt_rel}.

\begin{definition}
  \label{def:vfcx_cbsem__ch2o_cbsem.stmt_rel}
  Relation \coqinline{stmt_rel} in \coderef{vfcx_cbsem__ch2o_cbsem.v} inductively defines the mapping between a \vfcx\ statement \coqinline{s\\_v} and CH2O core C statement \coqinline{s\\_h}:
\begin{lstlisting}[
    language=Coq,
    style=coqfootnotenumbered,
    linerange={389-395},
    firstnumber=389
  ]
Inductive stmt_rel: list string \rightarrowSPC vfcx.stmt \rightarrowSPC statements.stmt K \rightarrowSPC Prop :=
| stmt_rel_skip: \forallSPC xs,
    stmt_rel xs vfcx.StmtSkip skip
| stmt_rel_seq: \forallSPC xs s1\_vSPC s2\_vSPC s1\_hSPC s2\_h,
    stmt_rel xs s1\_vSPC s1\_hSPC \rightarrow
    stmt_rel xs s2\_vSPC s2\_hSPC \rightarrow
    stmt_rel xs (vfcx.StmtSeq s1\_vSPC s2\_v) (s1\_hSPC ;; s2\_h)
\end{lstlisting}
  The first interesting case is \coqinline{stmt_rel_let}, the assignment of a new variable. Since \vfcx\ does not allow uninitialized variables, we must immediately assign a value to it. Nevertheless, from the point of view of CH2O core C \emph{small step semantics}, there will still be a ``brief'' in-between step in the execution of the program in which the variable was declared but not assigned:
\begin{lstlisting}[
    language=Coq,
    style=coqfootnotenumbered,
    linerange={396-416},
    firstnumber=396
  ]
| stmt_rel_let: \forallSPC xs x e\_vSPC s\_vSPC e\_hSPC s\_h,
    x \notinSPC xs \rightarrow
    expr_rel_int (x :: xs) e\_vSPC e\_hSPC \rightarrow
    stmt_rel (x :: xs) s\_vSPC s\_hSPC \rightarrow
    stmt_rel xs
      (vfcx.StmtLet x e\_vSPC s\_v)
      (statements.SLocal sintT%BT (var 0 ::= cast{sintT%T} e\_hSPC ;; s\_h))
| stmt_rel_assign: \forallSPC xs x e\_vSPC i e\_h,
    x \inSPC xs \rightarrow
    xs !! i = Some x \rightarrow
    expr_rel_int xs e\_vSPC e\_hSPC \rightarrow
    stmt_rel xs
      (vfcx.StmtExpr (vfcx.ExprAssign (vfcx.ExprVar x) e\_v))
      (! (cast{voidT%T} (var i ::= e\_h)))
| stmt_rel_if: \forallSPC xs e\_vSPC s1\_vSPC s2\_vSPC e\_hSPC s1\_hSPC s2\_h,
    expr_rel_bool xs e\_vSPC e\_hSPC \rightarrow
    stmt_rel xs s1\_vSPC s1\_hSPC \rightarrow
    stmt_rel xs s2\_vSPC s2\_hSPC \rightarrow
    stmt_rel xs
      (vfcx.StmtIf e\_vSPC s1\_vSPC s2\_v)
      (if{e\_h} s1\_hSPC else s2\_h)
\end{lstlisting}
  The other interesting case is \coqinline{stmt_rel_while} handling the C while loop, which in CH2O core C needs to be coded using core C's \coqinline{loop}, \coqinline{throw} and \coqinline{catch} mechanism:
\begin{lstlisting}[
    language=Coq,
    style=coqfootnotenumbered,
    linerange={417-428},
    firstnumber=417
  ]
| stmt_rel_while: \forallSPC xs e\_vSPC i s\_vSPC e\_hSPC s\_h,
    expr_rel_bool xs e\_vSPC e\_hSPC \rightarrow
    stmt_rel xs s\_vSPC s\_hSPC \rightarrow
    stmt_rel xs
      (vfcx.StmtWhile e\_vSPC i s\_v)
      (catch (loop ((if{e\_h} skip else throw 0) ;; catch s\_h)))
| stmt_rel_ret: \forallSPC xs e\_vSPC e\_h,
    expr_rel_int xs e\_vSPC e\_hSPC \rightarrow
    stmt_rel xs
      (vfcx.StmtRet e\_v)
      (ret (cast{sintT%T} e\_h))
.
\end{lstlisting}
\end{definition}
Relation \coqinline{stmt_rel} can be understood as a constraint, limiting application of our overall approach to those programs for which the relation can be proven.

\begin{theorem}[Soundness of \cbsem\ with respect to CH2O big step semantics for functions]
  \label{thm:vfcx_cbsem__ch2o_cbsem.exec_func}
  From \coderef{vfcx_cbsem__ch2o_cbsem.v}:
\begin{lstlisting}[
    language=Coq,
    style=coqfootnotenumbered,
    linerange={739-745},
    firstnumber=739
  ]
Theorem exec_func: \forallSPC xs xs\_vSPC s\_vSPC q\_vSPC s\_h,
  xs \equiv\_pSPC elements xs\_vSPC \rightarrow
  stmt_rel xs s\_vSPC s\_hSPC \rightarrow
  vfcx_cbsem.exec_func_correct xs\_vSPC vfcx.ExprTrue s\_vSPC q\_vSPC \rightarrow
  \forallSPC (\sigma\_h: list (option Z)),
    Forall is_Some \sigma\_hSPC \rightarrow
    ((ch2o_cbsem.exec_func_correct (length xs) s\_h) \sigma\_h).
\end{lstlisting}
  It states that when correctness of a function has been proven by \cbsem\ (that is, by constructing a derivation using Definition \ref{def:vfcx_cbsem.exec_func_correct}), we may conclude correctness of a function as stated in Definition \ref{def:ch2o_cbsem.exec_func_correct}.

  The conditions for this to hold are that the functions take the same arguments (expressed using STDPP's permutational equivalence \coqinline{\\equiv\\_p}) and the demand that the \vfcx\ and CH2O core C programs are related in the sense of Definition \ref{def:vfcx_cbsem__ch2o_cbsem.stmt_rel}.
\end{theorem}

\begin{theorem}[Soundness of \cbsem\ with respect to CH2O big step semantics for programs]
  \label{thm:vfcx_cbsem__ch2o_cbsem.exec_prog}
  Also here, the soundness for entire programs follows trivially from the soundness for functions:

\begin{lstlisting}[
    language=Coq,
    style=coqfootnotenumbered,
    linerange={770-773},
    firstnumber=770
  ]
Lemma exec_prog: \forallSPC s\_hSPC s\_v,
  stmt_rel [] s\_vSPC s\_hSPC \rightarrow
  vfcx_cbsem.exec_prog_correct vfcx.ExprTrue s\_vSPC vfcx.ExprTrue \rightarrow
  ch2o_cbsem.exec_prog_correct s\_h.
\end{lstlisting}
\end{theorem}

\subsection{Soundness of CH2O big step semantics with regards to CH2O small step semantics}

Finally we arrive at proving soundness with respect to CH2O's core C own small step semantics. We use CH2O's \emph{$\rho$-extended small step relation} \coqinline{rcstep} (Definition 8.6.3 in \cite{krebbers-ch2o-phd-2015}) between states, which is parameterized by:
\begin{itemize}
  \item a \emph{type environment} \coqinline{\\Gamma} describing the types of \vfcinline{struct}/\vfcinline{union} fields and the types of functions (see Definition 3.3.3 in \cite{krebbers-ch2o-phd-2015});
  \item a \emph{function environment} (or simply, a core C \emph{program}) \coqinline{\\delta} mapping function names onto function bodies (see Definition 6.2.5 of \cite{krebbers-ch2o-phd-2015});
  \item a \emph{stack} \coqinline{\\rho} extending the local stack (see Definition 6.1.10 in \cite{krebbers-ch2o-phd-2015}). Each execution state has a local execution context, representing the statement or expression surrounding the present focus of execution; the extension with its small tweaks prevents future reductions from ``escaping'' into this surrounding execution context.
\end{itemize}
Full details of this semantics can be found in chapters 6 and 8 of \cite{krebbers-ch2o-phd-2015}. The CH2O code base uses notation \coqinline{\\Gamma \\ \\delta \\ \\rho \\vdash\\_s S1 \\Rightarrow S2} for a single step instance of \coqinline{rcstep} between two execution states \coqinline{S1} and \coqinline{S2}. The reflexive-transitive closure of the relation, describing a multiple step relation, simply adds a star to the notation (\coqinline{\\Rightarrow *}).

\begin{definition}
  \label{def:ch2o_rs.exec_func_correct}
  Correctness of a function in terms of CH2O core C small step semantics is defined in \coderef{ch2o_rs.v}:
\begin{lstlisting}[
    language=Coq,
    style=coqfootnotenumbered,
    linerange={14-23},
    firstnumber=14
  ]
Definition exec_func_correct \GammaSPC \deltaSPC f s := \forallSPC S,
  \checkmarkSPC \GammaSPC \rightarrow
  \checkmark{\Gamma,'{\emptyset}} \deltaSPC \rightarrow
  (\Gamma, '{\emptyset}, []) \vdashSPC s : (true, Some sintT%T) \rightarrow
  \Gamma\ \delta\ [] \vdash\_sSPC State [CParams f []] (Stmt \searrowSPC s) \emptysetSPC \Rightarrow* S \rightarrow
  (
    red (rcstep \GammaSPC \deltaSPC []) S
    \lor
    \existsSPC m z, S = State [] (Return f (intV{sintT} z)) m
  ).
\end{lstlisting}
  The first two premisses in this definition relate to the \emph{validity} or well-formedness of the type and function environments (see Definitions 3.3.6 and 6.2.9 in \cite{krebbers-ch2o-phd-2015}, respectively). The third premisse states that the function will either diverge or terminate with a \vfcinline{return} statement returning an \vfcinline{int}. The fourth premisse states that there is a multi-step reduction starting from some initial state (\emph{entering} the function body \coqinline{s} with no parameters) and ending in \emph{some} state \coqinline{S}. The correctness definition then requires that:
  \begin{itemize}
    \item either we can take another step, starting from this state \coqinline{S} (\emph{progress});
    \item or \coqinline{S} must be a return state with some integer value \coqinline{z} and memory state \coqinline{m}.
  \end{itemize}
\end{definition}
\begin{definition}
  \label{def:ch2o_rs.exec_prog_correct}
  \coderef{ch2o_rs.v} concludes with a notion of program correctness building on top of function correctness:
\begin{lstlisting}[
    language=Coq,
    style=coqfootnotenumbered,
    linerange={35-35},
    firstnumber=35
  ]
Definition prog_correct \GammaSPC \deltaSPC s := exec_func_correct \GammaSPC \deltaSPC "main" s.
\end{lstlisting}
\end{definition}

\begin{theorem}[Soundness of CH2O big step semantics with respect to CH2O core C small step semantics for functions]
  \label{thm:ch2o_cbsem__ch2o_rs.exec_func}
  We can now prove soundness of function correctness from Definition \ref{def:ch2o_cbsem.exec_func_correct} with respect to Definition \ref{def:ch2o_rs.exec_func_correct}. From \coderef{ch2o_cbsem__ch2o_rs.v}:
\begin{lstlisting}[
    language=Coq,
    style=coqfootnotenumbered,
    linerange={1154-1156},
    firstnumber=1154
  ]
Lemma exec_func: \forallSPC f s,
  (\forallSPC \sigma\_h, ch2o_cbsem.exec_func_correct 0 s \sigma\_h) \rightarrow
  ch2o_rs.exec_func_correct \GammaSPC \deltaSPC f s.
\end{lstlisting}
\end{theorem}
The proof for Theorem \ref{thm:ch2o_cbsem__ch2o_rs.exec_func} obviously deals with the termination and divergence cases separately. In both cases, we make use of a separate \emph{axiomatic semantics} defined for CH2O core C in chapter 8 of \cite{krebbers-ch2o-phd-2015}. An adequacy theorem found in CH2O's Coq library (\coqinline{ax_stmt_adequate}, related to the adequacy theorem 8.6.12 from the thesis \cite{krebbers-ch2o-phd-2015}) links correctness of judgments of CH2O's axiomatic semantics to the outcome state \coqinline{S'} of a multi-step reduction. By working in the axiomatic semantics, we can represent the store defined in \ref{def:ch2o_store.store} as an assertion in the separation logic of that axiomatic semantics.

\begin{definition}
  \label{def:assert_stack}
  Separation logic predicate \coqinline{assert_stack} maps a \coqinline{store} of temporaries to a separating conjunction of heap-allocated locals. From \coderef{ch2o_store.v}:
\begin{lstlisting}[
    language=Coq,
    style=coqfootnotenumbered,
    linerange={46-51},
    firstnumber=46
  ]
Fixpoint assert_stack \sigma: assert K :=
  match \sigmaSPC with
  | [] => emp
  | mv :: \sigmaSPC =>
      assert_var_points_to 0 mv \starSPC assert_stack \sigmaSPC \uparrow
  end.
\end{lstlisting}
\end{definition}
Using CH2O core C's axiomatic semantics in this way allows us to ``bypass'' a lot of bookkeeping that would arise if we proved soundness of CH2O big steps semantics with respect to the small step semantics \emph{directly}. The only inference rule that appeared to be lacking in the axiomatic semantics for CH2O core semantics, at least for our purposes, was a rule for \emph{loop unrolling}. Given that CH2O is a shallow embedding, we added and proved this inference rule as a lemma.
\begin{lemma}[Loop unrolling in CH2O core C axiomatic semantics]
  From \coderef{ch2o_extensions.v}:
\begin{lstlisting}[
    language=Coq,
    style=coqfootnotenumbered,
    linerange={140-143},
    firstnumber=140
  ]
Lemma ax_loop_unroll R J T C P P' Q s:
  \Gamma\ \delta\ R\ J\ T\ C \models\_sSPC {{ P }} s {{ P' }} \rightarrow
  \Gamma\ \delta\ R\ J\ T\ C \models\_sSPC {{ P' }} loop s {{ Q }} \rightarrow
  \Gamma\ \delta\ R\ J\ T\ C \models\_sSPC {{ P }} loop s {{ Q }}.
\end{lstlisting}
  (\coqinline{R}, \coqinline{J}, \coqinline{T} and \coqinline{C} are environments related to non-local controls \vfcinline{return}, \vfcinline{goto}, \coqinline{throw} and \vfcinline{switch}, respectively.)
\end{lemma}

\begin{theorem}[Soundness of CH2O big step semantics with respect to CH2O core C small step semantics for programs]
  \label{thm:ch2o_cbsem__ch2o_rs.exec_prog}
  Proving soundness for CH2O big step programs (Definition \ref{def:ch2o_cbsem.exec_prog_correct}) with respect to Definition \ref{def:ch2o_rs.exec_prog_correct} is trivial. From \coderef{ch2o_cbsem__ch2o_rs.v}:
\begin{lstlisting}[
    language=Coq,
    style=coqfootnotenumbered,
    linerange={1171-1173},
    firstnumber=1171
  ]
Lemma exec_prog: \forallSPC \GammaSPC \deltaSPC s\_h,
  ch2o_cbsem.exec_prog_correct s\_hSPC \rightarrow
  ch2o_rs.prog_correct \GammaSPC \deltaSPC s\_h.
\end{lstlisting}
\end{theorem}


\section{Future work}
\label{section:future_work}

While the present subset of C features supported is sufficient to demonstrate the validity of our approach, the most important task ahead is to add extra C language features, such as pointers, \vfcinline{struct} types, dynamic memory allocation and function calls. We also need to translate supporting VeriFast constructs such as predicates, lemmas and fixpoint functions, inductive types and ghost statements such as \vfcinline{open}, \vfcinline{close}, \vfcinline{assert} and \vfcinline{leak}. Adding support for concurrent programs would be a very powerful feature, but would be complicated by the fact that we rely on big step semantics and that both CH2O and CompCert semantics themselves are single-threaded. It may be useful to port CH2O itself to make use of the Iris framework \cite{jung-iris-jfp-2018}.

Extending the set of supported features will involve more than expanding the semantics and corresponding soundness proofs. Currently, when proving the symbolic execution proposition in Coq, we can discharge many of our proof goals using Coq tactic \coqinline{lia}. As the set of supported types and features grows and programs become more complex, this will not work anymore. At that point we will need to instrument the SMT solver and use its internal state to export a proof script for each proof obligation.

Another goal is to retain our support for targeting CompCert. While this report describes how we replaced CompCert with CH2O because of the sophisticated memory model of the latter, soundness of symbolic execution with regards to CompCert remains very valuable, because it allows us to extend our correctness guarantees to the assembly code generated by CompCert. Luckily, we can have it all at once: given that CompCert C, like CH2O, is C11-compliant, we can recover this earlier result by extending the chain of soundness proofs with one more step, proving CH2O small step semantics sound with regards to CompCert C.

Finally, we must also look at other work aiming to obtain correctness proofs for C programs that can be machine-checked by a generic proof assistant. Especially relevant is the \emph{Verified Software Toolchain} (VST) \cite{appel-sf5-2021} which is very mature in terms of supported language features and allows expression of complex specifications, but is less ergonomic and efficient than VeriFast and only provides guarantees with respect to CompCert C, and not with respect to other C11-compliant compilers. Other relevant projects include \emph{Boogie} \cite{parthasarathy-boogie-cav-2021}, \emph{Refined C} \cite{sammler-refined-c-pldi-2021}, \emph{BRiCk} \cite{brick} and \emph{CN} \cite{pulte-popl-2023}. Once our feature set is more complete, we can perform a more in-depth comparison to these projects.


\bibliographystyle{acm}
\bibliography{index}


\end{document}